# Design Challenges in High-Current Pulsed Striplines

H. Pfeffer, M. Davidson, N. Curfman, T. Omark
Fermi National Accelerator Laboratory
Batavia, USA
pfeffer@fnal.gov, mattd@fnal.gov, ncurfman@fnal.gov, tomark@fnal.gov

*Abstract*—The Long Baseline Neutrino Facility (LBNF) will produce the world's most intense neutrino beam. Three series connected magnetic horns will require 5 kV, 300 kA, 800 μs pulses at a rate of 1.4 Hz to focus the beam. Connecting a single power supply to these focusing horns will require a low impedance connection measuring over 60 m in length. To meet the challenging requirements of connecting the horns to the power supply, this connection is engineered as a nine-conductor, high-current pulsed stripline. It must pass through a harsh radiation environment, be passively cooled, and have an operational lifetime of at least 30 years. This paper discusses mechanical and electrical considerations such as high-voltage holdoff, clamped joint performance, and Lorentz force mitigations in order to meet the specified requirements. The results of tests and experiments on several prototypes of key design features will be presented and discussed.

## I. Introduction

To deliver energy to the horns, a parallel plate stripline was developed to create a low impedance path from the power supply. This is necessary as the power supply is located away from the horns to avoid operating in a high radiation environment [1], and thus the stripline must traverse from the power supply room through a target hall battlement and into the horn bunker as shown in Fig. 1.

The geometry and layout of the stripline is chosen such that the resistive losses and inductance are minimized while remaining relatively compact. To ease fabrication and installation, a joint design was developed and tested to clamp sections of the stripline together. The operation at peak currents of 300 kA, however, while simultaneously operating in a harsh radiation environment presents several design challenges. This paper discusses the arrangement and sizing of the conductors to minimize Lorentz forces and impacts on resistance and inductance. The design of the joint, as well as thermal and high-voltage considerations will be discussed. Prototypes of several stripline features were built and used to verify high-voltage performance and the impact of losses due to the joint. The salient results of those tests will be presented.

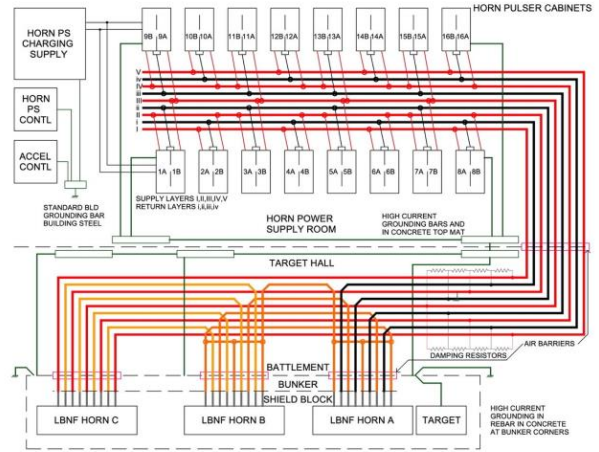

Fig. 1. Overview of stripline connection between power supply and the three horns.

## II. Design

### A. Conductor Arrangement

To minimize cost, inductance, and losses, the stripline is made up of nine 305 mm (12 in) wide, 9.53 mm (0.375 in) thick bus bars radiused to 1.91 mm (0.075 in) spaced apart at 12.7 mm (0.5 in) as shown in Fig. 2. Five of the nine plates are high-voltage (supply) and four are low-voltage (return). During a pulse, each of the inner seven conductors carry 75 kA, but only 37.5 kA flows through each outer conductor due to the coupled nature of time-varying electromagnetic fields. This is advantageous, as the outer plates are the only plates to experience any significant Lorentz forces. With the current halved, the force is reduced by a factor of four as seen in (1), where $I_0$ is the current flow through the outer plate, $w$ the width of the plate, and $\mu_0$ the permeability of free space. The forces are independent of plate spacing and thickness.

$$F = 0.5\, \mu_0 \frac{I_0^2}{w} \quad (1)$$

An Opera 2D simulation calculated a force of 2.7 kN/m with 37.6 kA in the outer plates.

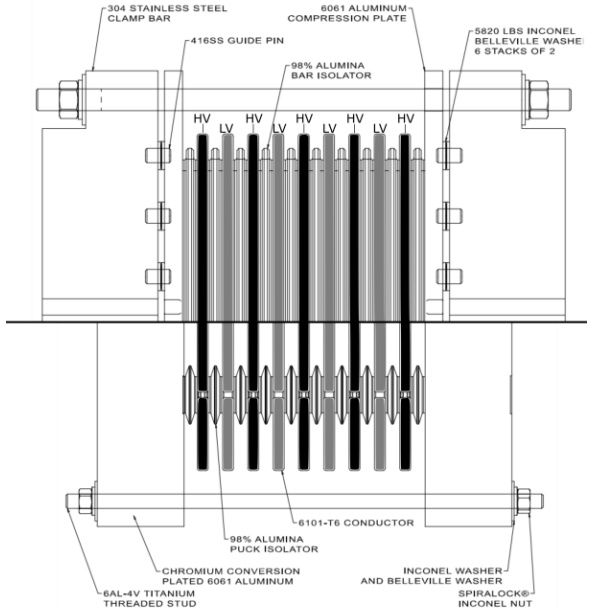

*Fig. 2. Cross-section view of stripline joint clamp (top) with flat-bar insulators and standard clamp (bottom) with puck insulators.*

It is desirable for the resistance and inductance of the stripline to be minimized not only to minimize losses but to limit the required operating voltage ($V_{op}$) as determined by (2).

$$V_{op} = L\frac{dI}{dt} + RI \qquad (2)$$

Where *L*, *R*, and *I* are the effective inductance, resistance, and current of the series stripline and horns. The stripline plate spacing is optimized to obtain the lowest inductance while not exceeding withstand voltage requirements. The thickness of each conductor was chosen with regard to a given skin depth of 3.4 mm and the fact that both sides of each internal conductor carry current.

The stripline electrical parameters were obtained using Opera 2D to determine the per-unit length energy storage in the magnetic and electric fields. Using the provided dimensions, current distribution, and operating frequency of 625 Hz, the per-unit inductance and capacitance are calculated to be 7.83 nH/m and 1.764 nF/m respectively. The magnetic field simulation also provides a power per-unit length based on the properties of the 6101-T6 aluminum material, from which the per-unit resistance of the stripline is found to be 6.72 µΩ/m. The comparison of a 65 m stripline to the resistance and inductance of the three horns can be seen in Table I.

### B. Joint Design

Previous striplines at Fermilab have only sparingly used clamped joints, and if a 90° turn of the stripline could not be achieved by bending the conductors, two conductors were friction stir welded (FSW) together. Due to a recent decrease in capable and available vendors, FSW joints are used sparingly. Instead, the ends of conductors to be joined are machined, plated, and clamped together with heavy-duty bar clamps providing 11.7 MPa (1700 lbf/in$^2$) of clamping pressure. The joint area has been sized to keep the current density under 200 A-rms/in$^2$ as required for plated joints [2] while optimizing joint overlap for streamline effects [3] and joint pressure to minimize joint resistance.

TABLE I. LOAD PARAMETERS

| Inductance and Resistance Values for the Stripline and Horn | | |
|---|---|---|
| *Load Element* | *Inductance (µH)* | *Resistance (mΩ)* |
| Horn A | 0.745 | 0.215 |
| Horn B | 1.101 | 0.078 |
| Horn C | 0.393 | 0.027 |
| Stripline (65 m) | 0.508 | 0.437 |

Traditionally, silver plating has been used in stripline connections for its high conductivity, but its cost and durability are not as favorable as nickel plating. The three-layer right-angle joint prototype was designed with two silver-plated joints and one nickel-plated joint to determine what effect switching to nickel-plated conductors has on joint resistance.

### C. Thermal Considerations

Radiation concerns dictate the stripline be cooled by natural convection. The stripline is surrounded by a ventilated enclosure to limit temperature rise from Joule heating to 30 °C with a calculated 350 W/m power loss. Expansion loops similar to those used in industrial piping must be placed strategically throughout the structure to allow the conductors to expand and contract with an additional 30 °C ambient temperature fluctuation.

### D. Material Limitations

The stripline passes through nitrogen-pressurized radiation environments that are difficult and time consuming for personnel to access. To ensure reliability, materials have been carefully selected based on previous experience with similar beamlines. These harsh environments necessitate the use of aluminum conductors and a mix of stainless steel, Inconel, and titanium hardware. Corrosive compounds formed when radiation interacts with nitrogen reduces the lifetime of copper bus bars, and radiation-induced hydrogen embrittlement limits the service life of high-carbon steel fasteners.

### E. High-Voltage Considerations

High-voltage considerations of the stripline mainly come in the form of chosen geometry and spacing along with the design of insulators in the clamped regions. The environment in which this operates is both in air and a highly ionized nitrogen environment. Fermi's design standards are to limit the max field strength to 1.42 kV/mm (36 kV/in), about 50% the uniform field limit of air. Prior experience within highly ionized environments dictates limiting the field strength further to 0.98 kV/mm (25 kV/in). The leakage path for insulators in this environment is designed for a minimum of 5.08 mm/kV (0.2 in/kV) and therefore, should achieve an effective path length of at least 25.4 mm (1 in). These figures are markedly conservative but allow for robustness against any surface imperfections and a higher degree of reliability, which is needed as repairs are extremely difficult and insulators are not removable once installed.

It is notable that since the mechanical structure is grounded, it may be more desirable to have the return lines on the outside to minimize interactions with outer structures. However, due to various positioning constraints, design of the other conductor structures was made substantially easier by having the outer plates at high voltage.

In places of clamping, insulators are used to maintain both mechanical and electrical integrity of the stripline. The insulator's material, geometry, and ease of manufacture are the main considerations. Alumina ($Al_2O_3$) ($\varepsilon$=9.9) of greater than 98% purity was chosen due to its relatively low cost, ease of machining, and prior proven use in previous stripline designs from NuMI and MiniBooNE [4]. Additionally, ceramics tend to be favorable for a clamped style of design due to their naturally high compressive strength and resistance to surface damage from leakage currents [5].

Two profiles are used depending on which clamp is used as can be seen in Fig. 2: small ceramic pucks, used in the majority of the stripline at standard clamp locations, and larger flat-bar insulators, used where clamps are electrically joining two pieces of stripline together to provide an even clamping pressure. The puck and flat-bar insulators each have a single convolution: triangular and rectangular profiles with appropriately radiused or chamfered edges respectively. The differing profile shapes were a balance of manufacturability and electrical performance, simulated and optimized with Opera 2D. The leakage lengths of the puck and flat-bar insulator are 25.99 mm and 29.54 mm respectively.

### III. TESTING

#### A. High Voltage Testing

To verify the structures and geometries used for production, key design prototypes of the stripline were built to scale for partial discharge testing. To provide headroom and confidence that a flashover will not occur in operation, it was required that each prototype be free of corona discharge in excess of 100 pC up to 10 kVp (or 7.1 kV-rms). The test and measurement is performed with a commercial detector that outputs a high-voltage 60 Hz ac waveform. The results of the 60 Hz excitation can be extrapolated to higher frequencies of operation and the use of an ac waveform is advantageous in testing due to the ac nature of the voltage during a pulse.

In general, three tests for each prototype were performed that alternated potentials between plates and from the plates to the outer structure. In every case, the surrounding structure was at ground potential. The purpose of the tests is to not only emulate operating situations, but to isolate problematic regions.

As of this paper's writing, three different prototypes have been manufactured: a five-layer jumper prototype, a three-layer right-angle joint prototype, and a nine-layer prototype with connection tabs for 1/0 AWG lugs. The prototypes are full-scale but are only between 0.6–0.9 m in length as seen in Fig. 3. Additionally, the number of layers used varies based on the amount necessary to test a particular feature; which helped lower costs.

*1) Jumper Prototype Results*

The jumper prototype contains a tap which joins the outer/odd layers electrically. Since this tap is at high-voltage, the tap construction method was tested to ensure no issues. In general, 8.9-9.2 kV-rms was obtains before a 100 pC partial discharge signature was found. However, with the outer plates at the same potential as the surrounding structure, voltages up to 12 kV-rms could be obtained (the maximum withstand voltage was not investigated).

An additional series of tests were performed to test the robustness against rough handling or installation mishaps, as light scuffs, knicks, and or marks may go unnoticed once installed. After initial testing, scratches and tooling marks were implemented in egregious locations to degrade performance. Although the max withstand voltage was reduced, the prototype still passed requirements. Additionally, any imperfections can most likely be sanded or smoothed out to easily re-acquire the original withstand voltage, being weary of filing particulates.

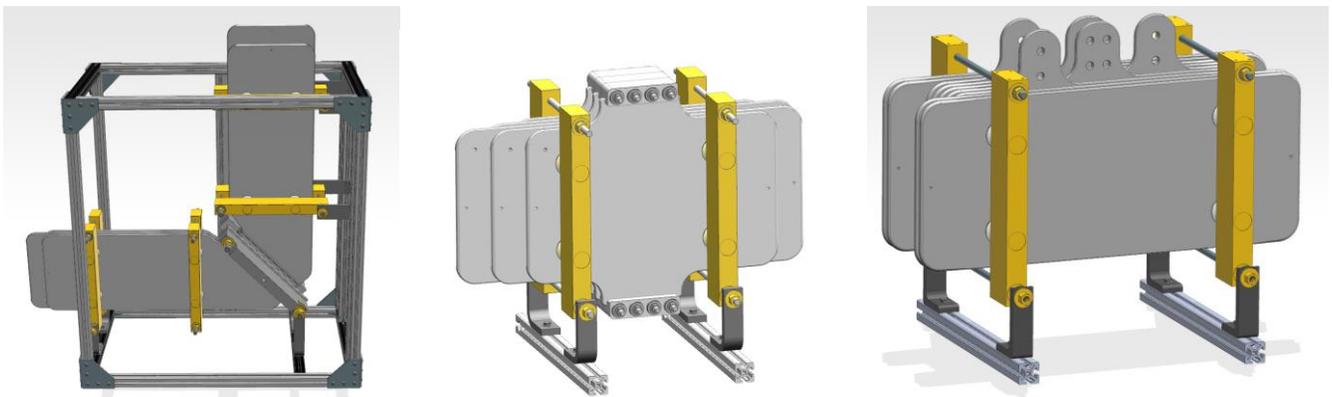

*Fig. 3. The three stripline prototypes: the joint prototype (left), the jumper prototype (middle), and the power supply connection prototype (right).*

*2) Joint Prototype Results*

The Joint prototype was fabricated to test the right angle joint clamp design, various plating materials, and the flat-bar insulators. Although slightly lower than the jumper prototype, the joint prototype still passed with an average of 8.5 kV-rms and similarly, an increase in withstand voltage to 9 kV-rms when the outer plates are grounded, although, not as prominent than that of the previous test.

A flashover event did occur during testing of this prototype unexpectedly around 10kV-rms, repeatably. It was found that debris, mostly small metal dust or shavings, had fallen on the flat-bar insulator from installing connections to a threaded hole. After a thorough cleaning, this event was unable to be reproduced. It is evident that large insulators with a relatively flat horizontal surface have the capability to collect foreign matter or contaminants. Therefore, care should be taken to the cleanliness of the insulators; dust may be a potential concern in some applications.

*3) Power Supply Connection Prototype Results*

The power supply connection prototype is the full 9-layer stripline with power supply connection tabs to test any issues with the connections to the wire lugs. With the outer plates at high potential, the average withstand voltage was a passing 9.8 kV-rms. This configuration was also tested till flashover, which occurred at 14.7 kV-rms. Instead of alternating the polarity, the middle plate was placed at high voltage and remaining eight plates at ground potential to test lugs and bolts but isolate any variables from the outer structure. The average withstand voltage was 11.9 kV-rms.

The flashover arcs occurred between the outer plates and support bars, particularly around one of the pucks. This and the higher withstand voltage with the outer plates grounded reinforces that higher voltage operation with this stripline would require more attention to this area. Additionally, the partial discharge waveform had very prominent behavior at the 90° phase resolved location, which suggests a potential geometry problem with the grounded hardware.

*B. Joint Resistance Testing*

Using nickel-plating rather than silver-plating in stripline joint connections is desirable due to the cost and durability of nickel. However, nickel has a lower conductivity than silver, which can lead to an increase in resistance and power loss at the joint. Resistance measurements at dc were made on the three-layer right-angle joint prototype to compare power loss at the joint of both silver-plating and nickel-plating.

To verify the clamping pressure was sufficient, resistance testing was performed as the clamp pressure was increased to verify the expected exponential decay in resistance. The clamping pressure is sufficient as the change in resistance approaches zero. The expected response for both silver and nickel were confirmed and the clamping pressure of 1900 lbf/in$^2$ agrees with recommendations from [3].

To measure dc resistances on the order of microhms, galvanic connection issues, temperature drift, and the absolute error of the measuring device need to be minimized. Therefore, voltage measurements at 40 A and then 0 A one second later were compared, which provided a repeatable measurement with a resolution of 0.5 µV with an Agilent 3458A. To evaluate the clamped joint plating, the above method was used to measure across a 50 mm (2 in) length of plate with no joint as compared to a similar length containing a joint as shown in Fig. 4. The difference in voltage between these two measurements can be attributed to the joint resistance. It was found that the nickel connection effectively adds the equivalent resistance of 75 mm (3 in) of stripline, while the silver connection effectively adds the equivalent resistance of 50 mm (2 in) of stripline.

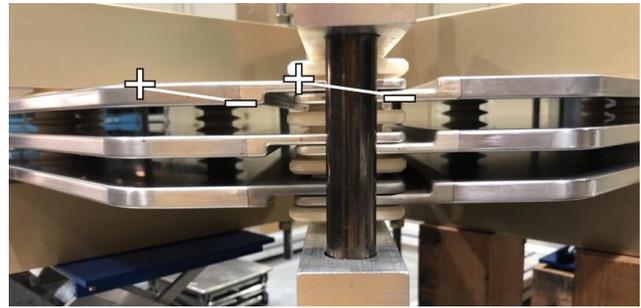

*Fig. 4. Measurement points for joint resistance testing.*

As previously stated, a per-unit resistance of 6.72 µΩ/m is expected in the 9-layer stripline. Using this per-unit resistance and the results from the joint resistance testing, an absolute resistance of the joint can be estimated. The 9-layer silver connection has an effective resistance of 0.34 µΩ while the 9-layer nickel connection has an effective resistance of 0.50 µΩ.

Nickel-plating will be used in the production stripline joint connections for its durability and cost over silver-plating. A dc resistance of 0.50 µΩ at a current of 7.17 kA-rms results in an additional loss of 25.7 W at the 9-layer joint, which is negligible in a stripline of this size.